\documentclass[apj,graphicx]{emulateapj}
\usepackage{apjfonts}

\newcommand\ngc{NGC~5548}
\newcommand\swi{{\it Swift}}

\makeatletter
\newcommand*{\rom}[1]{\expandafter\@slowromancap\romannumeral #1@}
\newcommand{\mbh}      {\ensuremath{M_{\mathrm{BH}}}}

\begin{document}

\title{{\it Swift}/UVOT grism monitoring of NGC 5548 in 2013: an attempt at \ion{Mg}{2} reverberation mapping}
\shortauthors{Cackett et al.}
\shorttitle{\ion{Mg}{2} in NGC 5548 with Swift}

\author{E.~M.~Cackett\altaffilmark{1}}
\author{K.~G\"{u}ltekin\altaffilmark{2}}
\author{M.~C.~Bentz\altaffilmark{3}}
\author{M.~M.~Fausnaugh\altaffilmark{4}}
\author{B.~M. Peterson\altaffilmark{4, 5}}
\author{J.~Troyer\altaffilmark{1}}
\author{M. Vestergaard\altaffilmark{6, 7}}

\email{ecackett@wayne.edu}

\affil{\altaffilmark{1}Department of Physics \& Astronomy, Wayne State University, 666 W. Hancock St, Detroit, MI 48201, USA}
\affil{\altaffilmark{2}Department of Astronomy, University of Michigan, 1085 S. University Ave, Ann Arbor, MI 48109, USA}
\affil{\altaffilmark{3}Department of Physics \& Astronomy, Georgia State University, Atlanta, GA 30303, USA}
\affil{\altaffilmark{4}Department of Astronomy, The Ohio State University, 140 West 18th Ave, Columbus, OH 43210, USA}
\affil{\altaffilmark{5}Center for Cosmology and Astroparticle Physics, The Ohio State University, 191 West Woodruff Ave, Columbus, OH 43210, USA}
\affil{\altaffilmark{6}Dark Cosmology Centre, The Niels Bohr Institute, Juliane Maries Vej 30, DK-2100 Copenhagen \O, Denmark}
\affil{\altaffilmark{7}Steward Observatory, University of Arizona, 933 N Cherry Avenue, Tucson, AZ 85721, USA}

\begin{abstract} 
Reverberation-mapping-based scaling relations are often used to estimate the masses of black holes from single-epoch spectra of AGN.  While the radius--luminosity relation that is the basis of these scaling relations is determined using reverberation mapping of the H$\beta$ line in nearby AGN, the scaling relations are often extended to use other broad emission lines, such as \ion{Mg}{2}, in order to get black hole masses at higher redshifts when H$\beta$ is redshifted out of the optical waveband.  However, there is no radius-luminosity relation determined directly from \ion{Mg}{2}.  Here, we present an attempt to perform reverberation mapping using \ion{Mg}{2} in the well-studied nearby Seyfert 1, NGC~5548.  We used \swi\ to obtain UV grism spectra of NGC~5548 once every two days from April to September 2013.  Concurrent photometric UV monitoring with \swi\ provides a well determined continuum lightcurve that shows strong variability.  The \ion{Mg}{2} emission line, however, is not strongly correlated with the continuum variability, and there is no significant lag between the two.  We discuss these results in the context of using \ion{Mg}{2} scaling relations to estimate high-redshift black hole masses.
\end{abstract}

\section{Introduction}
The mass of supermassive black holes (SMBHs) at the centers of galaxies, and their evolution with time is an important part of the picture of galaxy formation and evolution, as strong correlations between galaxy and black hole properties suggest they are closely connected, e.g., through the $M$--$\sigma$ and $M$--$L$ relations \citep[e.g.,][]{gebhardtetal00a,magorrianetal98,gultekinetal09b, kormendyho13}.  Much effort over the last several decades has gone into determining the masses of SMBHs in nearby galaxies,  through methods such as those using stellar \citep[e.g.,][]{gultekin14} and gas dynamics \citep[e.g.,][]{walsh13}, or reverberation mapping \citep[e.g.,][]{petersonetal04,bentz09}.  This nearby sample of SMBHs with black hole mass measurements can then form the basis of scaling relations that can be applied to much more distant objects where direct methods are not possible.  One of the most powerful of these scaling relations is the radius-luminosity, $R$--$L$, relation which relates the size of the broad emission line region (BLR) in an active galactic nucleus (AGN), as measured by reverberation mapping,  to the AGN luminosity.

The basic principle behind reverberation mapping is simple.  In AGNs, large amounts of photoionized gas moves under the influence of the central SMBH's gravity, allowing a direct measurement of the velocity dispersion in the BLR.  To determine a mass, the radius of the BLR from the SMBH is needed, which can be obtained using the reverberation mapping technique \citep{blandmckee82,peterson93,peterson14}.  In this method, the observed time lag, $\tau$, between an emission line lightcurve (typically H$\beta$) and the optical continuum lightcurve is interpreted as the light-travel time from the continuum emitting region close to the SMBH and the BLR further out (where the continuum is reprocessed into line emission).  The lag thus gives the emissivity-weighted average radius of the BLR, $R = \tau c$.  Combining some measure of the width of the emission line used and the radius leads to a mass measurement via the virial theorem.  In this way, reverberation mapping has successfully  determined the masses of $\sim$60 supermassive BHs in AGNs \citep[e.g.,][]{petersonetal04,bentz09,barth11,pancoast14,barth15,bentzkatz15}.

Such mass measurements have allowed the determination of the $R$--$L$ scaling relation \citep{kaspi00,bentz_rl_06,bentz_rl_09, bentz13}, which is based on the time lag (and hence radius) obtained using the H$\beta$ broad emission line.  The $R$--$L$ relation leads to a mass estimate from a single-epoch AGN spectrum --- a measurement of the AGN luminosity gives the BLR radius from the $R$--$L$ relation, and the broad emission line width can be used to determine the line of sight velocity dispersion.  This therefore allows estimates of black hole masses in large samples of galaxies \citep[e.g.][]{vestergaard02,mclure02,vestergaard06,vestergaard09}.  The $R$--$L$ relation has also successfully been used to discover low-mass ($\mbh < 2\times10^{6}$ M$_\odot$) AGN \citep{gh07, baldassare15}.

The $R$--$L$ relationship is well-established for H$\beta$, which is easily observed with ground-based observations in the case of local AGNs. However, as $z$ increases, H$\beta$ is shifted to IR wavelengths that become less accessible. At higher redshifts, it is thus desirable to use instead strong rest-frame UV lines, such as \ion{C}{4}\,$\lambda1549$ and \ion{Mg}{2}\,$\lambda2798$. Unfortunately, the $R$--$L$ relationship is poorly established for \ion{C}{4} and is heavily dependent on one provisional measurement of a high-luminosity quasar  \citep{kaspi07}. However, gravitational microlensing has allowed measurements of the size of the high-ionization BLR in gravitationally lensed systems, and these data support an $R$--$L$ relationship that is parallel to that of H$\beta$ \citep{guerras13}. The situation is even worse for \ion{Mg}{2} as there are only two reliable  \ion{Mg}{2} lags that have been measured, NGC 3783 \citep{reichert94} and NGC 4151, for which there are two independent measurements \citep{metzroth06}. Thus far, the best that can be done in the absence of suitable \ion{C}{4} and  \ion{Mg}{2} reverberation measurements is to assume these lines have $R$--$L$ relationships that are parallel to that for H$\beta$ and assume that all lines yield the same black hole mass, so that the quantity ${\rm VP} = R \Delta V^2/G$, where $\Delta V$ is the line width, is constant \citep{onken08,mcgill08, rafiee11,shen12,park13}.  Note that in all cases where it has been testable, VP is found to be consistent between different emission lines \citep{peterson99,peterson00,kollatschny03,petersonetal04,bentz10}.

Calibration of the black hole mass scale requires another step, usually written as $M_{\rm BH} = f {\rm VP}$, where $f$ is a dimensionless factor of order unity that depends on the BLR orientation and other generally poorly known parameters. The factor $f$ is thus expected to vary from system to system, but the consistency of the virial product for multiple lines in a given system suggests that $f$ is also approximately constant for a given system. With the most recent direct modeling of high-quality reverberation data, it is possible to uniquely model $f$ for a given system \citep{brewer11, pancoast14}, but at this time only a few such measurements have been made. What is usually done instead is to compute an ensemble average value for $\left< f \right>$ by using another estimate of the black hole mass, in practice from the $M$--$\sigma$ relationship.  
Two recent estimates from this method give $\left< f \right> = 4.31\pm1.05$ \citep{grier13} and $\left< f \right> = 4.47\pm1.24$ \citep{woo15}.  Interestingly, these are consistent with the average of the 5 individual $f$ values determined from direct modeling by \citet{pancoast14} (see that paper for a detailed discussion comparing $f$ values from the two independent and separate approaches).

There has been debate about the reliability of using \ion{C}{4} and \ion{Mg}{2} for black hole mass estimates.  For instance, \ion{Mg}{2} is systematically narrower than H$\beta$ \citep{wang09,marziani13}, but may be more reliable than \ion{C}{4} \citep{trakhtenbrot12}.  However, other studies have shown no net bias in using these rest-frame UV lines \citep{greene10,rafiee11}, and good agreement with H$\beta$-based masses \citep{assef11}.  Furthermore, the discrepancies and scatter in the relations can be a result of the non-variable component of line profile \citep{denney12} and low signal-to-noise ratio spectra that do not allow for an accurate characterization of the line profile \citep{denney13}.  Even so, measuring reverberation of \ion{Mg}{2} remains an important goal in order to further validate these scaling routines.  Thus far, there has only been one \ion{Mg}{2} mass determined \citep[NGC 4151,][]{metzroth06} using archival {\it IUE} data, which gives a value consistent with the H$\beta$ mass \citep{bentz06}.

 As discussed above, there have only been two objects where a \ion{Mg}{2} lag has been successfully recovered \citep{reichert94, metzroth06}.  It is useful to discuss other cases where reverberation of \ion{Mg}{2} has been looked for, and where the variability of \ion{Mg}{2} has been studied.  One of the more intense monitoring campaigns that looked for \ion{Mg}{2} variability was the 1989 {\it IUE} campaign of NGC~5548 \citep{clavel91}.  While there was significant variability in the continuum (ratio of maximum to minmum flux $\sim$4.5) and high-ionization lines, \ion{Mg}{2} was the least variable with a maximum to minimum flux ratio of $\sim$1.3. \citet{clavel91} attempted to measure a lag but found it was not well constrained, with $\tau = 34 - 72$ days.  During the monitoring campaign of NGC 3783, even though \citet{reichert94} were successfully able to measure a lag, \ion{Mg}{2} is again the least variable emission line, and the authors worry about whether variable \ion{Fe}{2} contributes to the \ion{Mg}{2} variability.  No other studies have had the intense monitoring required to attempt \ion{Mg}{2} reverberation, however, there are a number of studies looking at \ion{Mg}{2} variability from fewer observations.  Five {\it HST} observations of NGC~3516 also showed significant lack of variability in \ion{Mg}{2}, with the variability constrained to be less than 7\% even though the UV continuum varied by a factor of 5 \citep{goad99a,goad99b}.  More recent observations of \ion{Mg}{2} variability in higher luminosity quasars have found mixed results.  For instance, \citet{woo08} found significant variability (8--17\% rms) in 4 of the 5 quasars studied over 1--1.5 year rest-frame timescales, and \citet{hryniewicz14} find a 25\% change in \ion{Mg}{2} flux in the quasar LBQS~211$-$4538 from observations 6 months apart.  On the other hand, little or no \ion{Mg}{2} variability is seen in studies of two other quasars \citep{trevese07,modzelewska14}.  Finally, a study of spectral variability of quasars in the SDSS Stripe 82 region found \ion{Mg}{2} variability to be weak, and less variable compared to Balmer emission lines \citep{kokubo14}.  Thus, there are a number of examples of low variability in \ion{Mg}{2}, with only a few exceptions, and very few attempts at \ion{Mg}{2} reverberation mapping.

The \swi/UVOT allows a route to doing UV reverberation mapping using the U grism onboard.  \swi's observing schedule is flexible, allowing for short (1 or 2 ks) daily monitoring of AGN.  The effective area of the U grism peaks at around the wavelength of \ion{Mg}{2}, and thus can, in principle, be used to perform direct \ion{Mg}{2} reverberation mapping.    In order to expand the number of AGN with \ion{Mg}{2} reverberation mapping, and as a first step towards an ultimate goal of determining a \ion{Mg}{2} $R$--$L$ relation, we undertook a long-term ($\sim$6 month) monitoring campaign of \ngc\ in 2013.  \ngc\ (an S0/a Seyfert galaxy at $z = 0.01718$) was chosen since it is the best-studied reverberation-mapped AGN to date, with many years of monitoring \citep[see, e.g.,][and references therein]{peterson02, bentz07, bentz10, derosa15, edelson15}, and also had previous {\it IUE} data \citep{clavel91} to allow for a feasibility study. 

Our \swi/UVOT U grism monitoring campaign of \ngc\ took place from 2013 April 1 to September 12 (PI: Cackett).  We have also supplemented the UV continuum lightcurve with concurrent \swi\ photometric monitoring that took place immediately before and during our grism campaign.  An analysis of the UV photometric lightcurves during this period has already been presented by both \citet{kaastra14} and \citet{mchardy14}.   Furthermore, the broadband spectral energy distribution of \ngc\ during this period (including \swi\ U grism spectra) has been examined by \citet{mehdipour15}.

The data reduction is described in Section~\ref{sec:data}, and our time series analysis is given in Section~\ref{sec:cc}.  We discuss our results and their implications in 
Section~\ref{sec:disc}.

\section{Data Reduction}\label{sec:data}

\subsection{UV grism data}
Grism observations were taken from 2013 April 1 to 2013 September 12 approximately once every two days (on average).  Each observation typically consists of a total of 2 ksec exposure time on the source.  However, this is usually performed as two separate $\sim$1 ksec exposures, taken within a couple of hours of each other.  The \swi\ target IDs of the grism observations are 91711, 91737 and 91739.  Of the total 82 grism observations, 7 (all target ID 91737) were taken at a roll angle where a nearby star was dispersed adjacent to the dispersed \ngc\ spectrum.  It was not possible to cleanly extract a spectrum from these observations that did not contain continuum emission from the star.  Excluding those 7 observations, we are left with a total of 75 grism observations of \ngc.  Of the remaining 75 observations, 7 exposures (target ID 91739) had a  roll angle that put a zeroth-order image of a star at approximately 2000\,\AA\ in the dispersed first order spectrum of \ngc.  These observations could not be used for the mean and rms spectrum, but we were still able to calculate line fluxes because the first order spectrum around the \ion{Mg}{2} line is not contaminated.

To perform the \swi/UVOT grism data analysis, we use Paul Kuin's {\sc UVOTPY} software version 2.0.3 \citep{kuin14}, which is designed specifically for analysis of \swi/UVOT grism data.  Details of the calibration of the \swi/UVOT grism and of the software are given in \citet{kuin15}.

We extract the grism spectrum using the {\tt uvotgetspec} tool, and default parameters for the width of the extraction regions for the source and background.  For the vast majority of observations, a short UVW2 image is taken during the same pointing.  Since with grism spectroscopy the wavelength scale depends on the location of the zeroth-order image, the software uses the UVW2 image in order to anchor the wavelength scale.  In a small number of pointings, no photometric image was taken, leading to a more poorly defined anchor position.  

The 1$\sigma$ wavelength accuracy of the UV grism in clocked mode is 9\,\AA\ \citep{kuin15}, thus, we use the \ion{Mg}{2} line itself in order to provide a better wavelength determination.  For each spectrum we find the centroid of the \ion{Mg}{2} using the wavelength range where the flux of the line is $\geq$70\% of the peak value.  We find a mean absolute wavelength shift of 9\,\AA\ when a photometric image has been taken \citep[consistent with the \swi\ calibration:][]{kuin15}, and 39\,\AA\ when no photometric image exists.

\begin{figure}
\centering
\includegraphics[width=8cm]{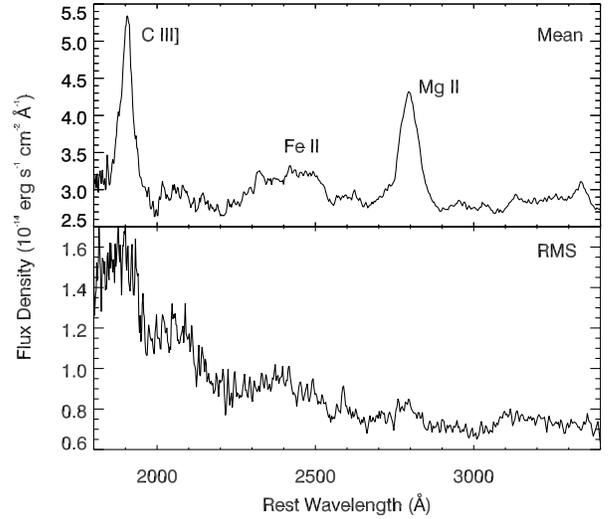}
\caption{{\it Top:} Mean UV grism spectrum of NGC 5548.  The \ion{C}{3}] and \ion{Mg}{2} emission lines are clearly visible.  The \ion{Fe}{2} emission line complex extends either side of the \ion{Mg}{2} line but is most prominent from 2200--2800\,\AA .  {\it Bottom:} Root mean square UV grism spectrum of NGC 5548.}
\label{fig:spec}
\end{figure}

Once the wavelength shifts have been applied, we calculate the mean and root mean square (rms) spectra, shown in Figure~\ref{fig:spec}.  All flux densities are given in the emitted frame, and are corrected for Galactic reddening assuming $E(B-V) = 0.0199$ \citep{schlegel98} and the dust reddening law of \citet{seaton79}.  We use \citet{seaton79} for easy comparison with previous work on NGC~5548 which also use this reddening law, though the choice does not make any difference in determining lags. The effective area of the UV grism drops off rapidly at about 1800\,\AA.  Furthermore, at wavelengths longer than about 3000\,\AA\, the second-order spectrum can overlap and contaminate the first-order spectrum.  We therefore concentrate on the spectrum between 1800--3400\,\AA.  The \ion{Mg}{2} line at a rest wavelength of 2798\,\AA\ can clearly be seen.  Other features in the spectrum include the \ion{C}{3}]\,$\lambda 1909$ semi-forbidden emission line, and the \ion{Fe}{2} emission line complex \citep[most prominent in the region 2200--2800\,\AA, see e.g.][]{baldwin04}.  \citet{mehdipour15} present a fit to the broadband spectral energy distribution of NGC~5548, which includes the mean UV grism spectrum, and the reader is referred to that paper for more details on individual components.  In the rms spectrum, the \ion{Mg}{2} can be identified, but it is not a strong feature, already indicating that it is not highly variable during our monitoring of \ngc. The mean continuum flux at 2670\,\AA\ is approximately the same as during the 1989 {\it IUE} observations presented by \citet{clavel91}.

The \ion{Mg}{2} line is well modeled by a single Gaussian.  Both a direct measurement and the best-fitting Gaussian give a FWHM = 68\,\AA.  However, this does not take into account the significant instrumental broadening.  Unfortunately, the line spread function is not accurately known at all wavelengths (N. P. M. Kuin, private communication).  The resolving power is given as R = 75 at 2600\,\AA\ \citep{kuin15}, which for the observed wavelength of \ion{Mg}{2} corresponds to $\Delta\lambda = 38$\,\AA.  Correcting for this broadening (assuming a Gaussian with FWHM = 38\,\AA) gives an intrinsic FWHM = 56.5\AA, or $\Delta v = 5960$~km~s$^{-1}$.  For comparison, the FWHM of $H\beta$ in \ngc\ has been seen to range from 3078~km~s$^{-1}$  \citep{peterson02} to 11177~km~s$^{-1}$ \citep{bentz10}, with line width anti-correlated with AGN luminosity.

\subsection{Photometric data}

\begin{figure*}
\centering
\includegraphics[width=15cm]{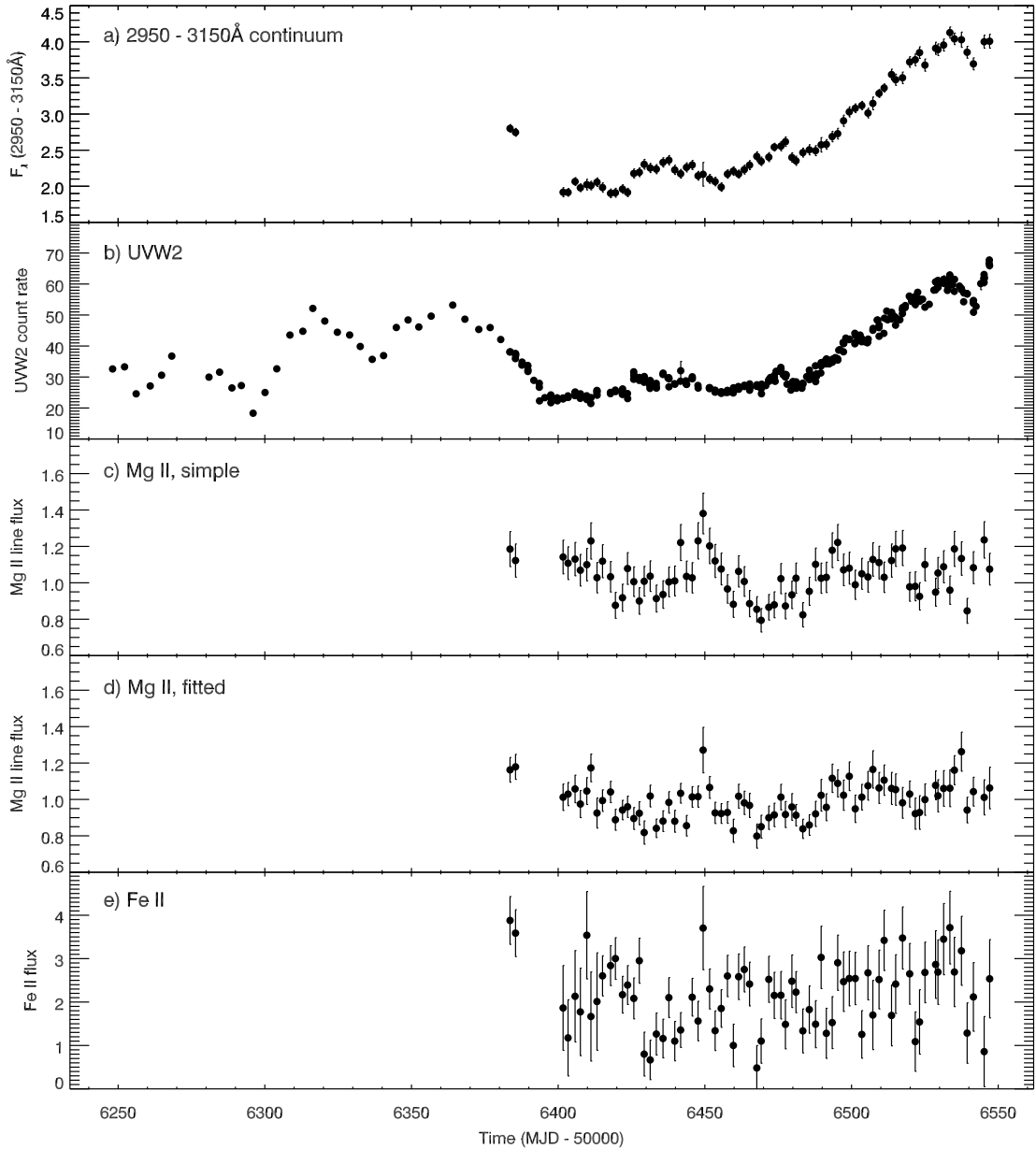}
\caption{{\it a)} Average flux density between 2950 and 3150\,\AA\ in units of $10^{-13}$ erg s$^{-1}$ cm$^{-2}$\,\AA$^{-1}$. {\it b)} \swi/UVW2 count rate in counts per second. {\it c)} Integrated \ion{Mg}{2} line flux in units of $10^{-12}$ erg s$^{-1}$ cm$^{-2}$ from the simple approach to determining the \ion{Mg}{2} line flux. {\it d)} Integrated \ion{Mg}{2} line flux in units of $10^{-12}$ erg s$^{-1}$ cm$^{-2}$ from the Fe template fitting approach to determining the \ion{Mg}{2} line flux. {\it e)} \ion{Fe}{2} flux integrated over the 2000 -- 3000\,\AA\ region in units of $10^{-12}$ erg s$^{-1}$ cm$^{-2}$ determined from Fe template fitting.}
\label{fig:lc}
\end{figure*}

As described above, \ngc\ was monitored intensely with \swi\ during 2013.  UVW2 was the filter most commonly used throughout the monitoring, thus, we use those data to determine the UV continuum lightcurve covering the period immediately before and throughout our grism monitoring.  In addition to the UVW2 observations associated with our grism observations (117 observations in total), we take advantage of other, shorter, photometric monitoring campaigns of \ngc\ taking place at the same time: target IDs 91404 (32 observations between 2012 November 17 and 2013 March 29, PI: McHardy),  91744 (50 observations, PI: Kaastra) and 30022 (17 observations).  Before 2013 April 1, photometric monitoring was approximately once every 4 days.  Beginning 2013 April 1 when our grism monitoring began, the cadence increased to once every 2 days, and from the end of May until mid-September monitoring occurred more frequently, sometimes with two observations per day.  In total we analyzed 216 observations between 2012 November 17 and 2013 September 12.

The UVW2 observations are often split into two separate pointings.  Rather than combine all pointings within a given day together, we analyze them separately in order to get the highest time cadence.  The UVW2 photometric lightcurve (along with other filters) during this period has already been presented by \citet{kaastra14}, \citet{mchardy14}, \citet{mehdipour15} and \citet{edelson15}. We perform photometry on \ngc\ using {\tt uvotsource} with a 5\arcsec\ circular source extraction region, and a 10\arcsec\ background region offset from the galaxy.  As described by \citet{mchardy14}, \citet{mehdipour15} and \citet{edelson15} a small number of observations were found to be anomalously low ($>$15\% lower) compared to the surrounding local mean, with the origin thought to be instrumental rather than intrinsic to the source.  We manually removed these ``drop-outs'' from the lightcurve (see \citealt{mchardy14}, \citealt{mehdipour15} and \citealt{edelson15} for more detailed discussion).  The UVW2 photometric lightcurve is shown in panel b) of Figure~\ref{fig:lc}.

\subsection{UV continuum and \ion{Mg}{2} lightcurves}

We use the individual grism spectra to determine the UV continuum flux and \ion{Mg}{2} lightcurves.  For the UV continuum lightcurve, we calculate the mean flux density from 2950 to 3150\,\AA.  This lightcurve is also shown in panel a) of Figure~\ref{fig:lc}.  As can be seen, the 2950--3150\,\AA\ lightcurve is strongly correlated with the UVW2 lightcurve.  The UVW2 bandpass peaks at a shorter wavelength at about 2100\,\AA\, and the effective area drops off significantly by 3000\,\AA. Figure~\ref{fig:fluxflux} shows the UVW2 count rate versus the 2950--3150\,\AA\ continuum flux, demonstrating the strong correlation between the two.  The UVW2 is highly variable with a variability amplitude \citep{vaughan03} of $F_{\rm var} = 0.33$.

To determine the \ion{Mg}{2} line flux we take two approaches.  First we take a simple approach involving defining the continuum in regions either side of the line, as is typically done in AGN reverberation studies, since the goal is to capture the line flux {\em variations} in a model-independent fashion, rather than to capture all the line flux.  Second, we perform multi-component spectral fitting to try and separate the contribution of the \ion{Fe}{2} complex from the \ion{Mg}{2} line and continuum.

\subsubsection{\ion{Mg}{2} line flux: simple approach}

\begin{figure}
\centering
\includegraphics[width=8cm]{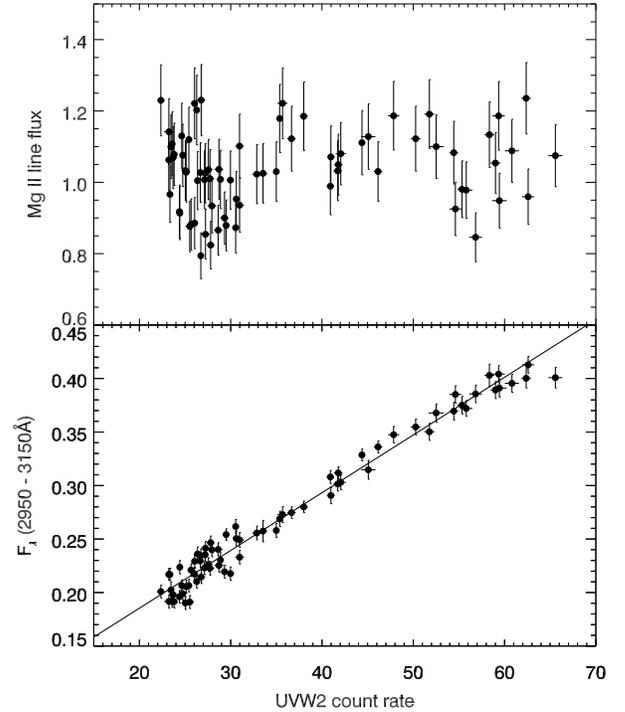}
\caption{{\it Top:} \ion{Mg}{2} line flux ($10^{-12}$ erg s$^{-1}$ cm$^{-2}$) from the simple method versus UVW2 count rate (c/s).  No correlation is apparent. {\it Bottom:} 2950--3150\,\AA\ flux density ($10^{-13}$ erg s$^{-1}$ cm$^{-2}$\,\AA$^{-1}$) versus UVW2 count rate.  There is a clear, strong correlation, with the best-fitting straight line shown.}
\label{fig:fluxflux}
\end{figure}

In this first approach, we determine the \ion{Mg}{2} integrated line flux by fitting the local continuum either side of the line.  We fit a straight line to the continuum including data in the ranges 2560--2695\,\AA\ and 2950--3150\,\AA.  We then integrate the line flux above the continuum from 2695--2900\,\AA.  The line lightcurve from this method is also shown in panel c) of Figure~\ref{fig:lc}.  We find that the uncertainties in line flux estimated directly from the uncertainties in individual flux bins appears to be overestimated, with mean fractional uncertainty in the line flux being 0.134.  We verified this overestimate by comparing flux differences between data points on timescales as short as 2 days.  Assuming those flux differences are stochastic, and that the fractional error is the same on all points, we get a fractional error of 0.081.  This is an upper limit on the flux uncertainties since there may be intrinsic flux variability on two-day timescales.  We adopt fractional errors of 0.081 on the \ion{Mg}{2} line flux measurements.  The lightcurve only has small amplitude variability, with $F_{\rm var} = 0.074$.  By eye, there does not appear to be a correlation between the continuum and \ion{Mg}{2} lightcurves, and this is also clear when looking at UVW2 count rate versus \ion{Mg}{2} line flux in Figure~\ref{fig:fluxflux}.

\begin{figure}
\centering
\includegraphics[width=8cm]{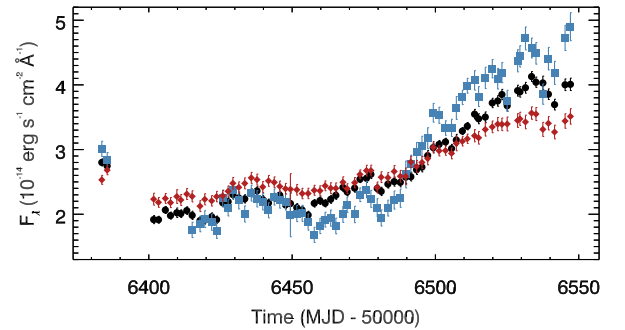}
\caption{Continuum lightcurves from the U grism in 2015--2215\,\AA\ (blue squares),  2950--3150\,\AA\ (black circles), and 4430--4625\,\AA\ (red diamonds).  They are all clearly correlated, but there is no significantly detected lag.  The amplitude of variability decreases with increasing wavelength, as expected for disk reverberation. \vspace{0.3cm}}
\label{fig:continuumlc}
\end{figure}

We also explore the continuum lightcurves at different wavelengths, calculating the mean flux density between 2015--2215\,\AA\ and also 4430--4625\,\AA.  We compare these lightcurves in Figure~\ref{fig:continuumlc}.  The lightcurves are clearly correlated, and there is an obvious decrease in variability amplitude with increasing wavelength.  This decrease in variability amplitude with wavelength can also be seen in the rms spectrum (Fig. \ref{fig:spec}).   The lightcurves indicate that the spectrum is bluer when brighter.

\subsubsection{Fe template fitting}

Strong \ion{Fe}{2} emission is visible in the spectrum, and thus to determine more robustly the \ion{Mg}{2} line flux we perform multi-component spectral fitting in order to decompose the \ion{Fe}{2}, \ion{Mg}{2} and continuum components.  A similar Fe template fitting approach has been used by \citet{barth13} to successfully recover optical \ion{Fe}{2} lags in two AGN.

For the \ion{Fe}{2} line complex we use the template model of \citet{vestergaard01}.  In addition to the Fe template, we also include a power-law continuum and a Gaussian to model the \ion{Mg}{2} line.  The model is then convolved with a Gaussian with FWHM = 38\,\AA\ to match the instrument resolution and fitted to the individual spectra in the 2000 -- 3000\,\AA\ region.

When fitting this model, we consistently found reduced-$\chi^2$ values significantly less than 1.0, once again indicating that the uncertainties in the flux from {\tt uvotsource} appear to overestimated.  We therefore scale the uncertainties by $\sqrt{\chi_\nu^2}$: the mean scale factor is 0.5.  

\begin{figure}
\centering
\includegraphics[width=8cm]{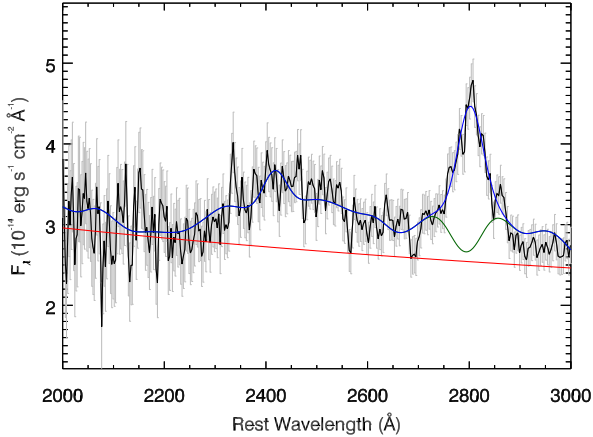}
\caption{{\it Swift} U grism spectrum of NGC 5548 from observation 00091711002.  The blue line shows the best-fitting spectral model, consisting of the power-law continuum (red), Fe template and a Gaussian to model the \ion{Mg}{2} line.  The green line shows the model minus the \ion{Mg}{2} line.}
\label{fig:fefitting}
\end{figure}

Figure~\ref{fig:fefitting} shows a spectral fit, using the first {\it Swift} observation in April 2013 (obsID: 00091711002) as an example.  The example demonstrates how the \ion{Fe}{2} emission overlaps with the \ion{Mg}{2} line.

\begin{figure}
\centering
\includegraphics[width=8cm]{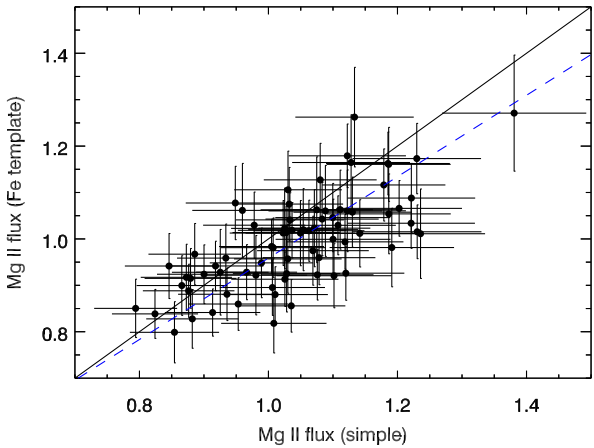}
\caption{A comparison of the \ion{Mg}{2} line flux ($10^{-12}$ erg s$^{-1}$ cm$^{-2}$) determined from the simple method and the Fe template fitting method.  The black solid line shows $x = y$, while the dashed blue line shows the best-fitting straight line.}
\label{fig:mg2comparison}
\end{figure}

There is good agreement between the line flux determined by the simple approach and the Fe template fitting method.  We show the \ion{Mg}{2} lightcurve from this method in panel d) of Figure \ref{fig:lc}, and a comparison of the line fluxes from both methods in Figure~\ref{fig:mg2comparison}.  The \ion{Mg}{2} flux is slightly lower when determined from the Fe template fitting method, because a small amount of Fe contributes to the total flux within the wavelength limits of the \ion{Mg}{2} emission line. The mean fractional uncertainty in the \ion{Mg}{2} line flux is 0.073 from this method, very similar to what we estimated for the simple method based on variability between adjacent  data points.

The integrated flux from the \ion{Fe}{2} line complex over the 2000 -- 3000\,\AA\ region is shown in panel e) of Figure~\ref{fig:lc}.  However, the \ion{Fe}{2} flux is poorly constrained from the spectral fitting, with a mean fractional uncertainty of 0.29.  The resulting lightcurve is noisy with no clear variability pattern.  Since the individual spectra are reasonably noisy, and the response of \ion{Fe}{2} is expected to be longer than that of H$\beta$ \citep{vestergaard05}, we also tried binning the spectra in time, by up to 5 epochs, to improve the lightcurve.  While this reduces scatter in the lightcurve, it does not show any clear correlated variability with the continuum lightcurve.

\section{Cross correlation Analysis}\label{sec:cc}
\begin{figure}
\centering
\includegraphics[width=8cm]{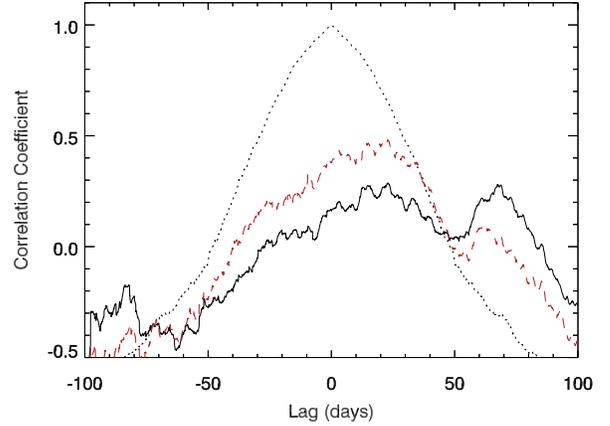}
\caption{The solid line shows the cross-correlation function (CCF) between the UVW2 and \ion{Mg}{2} lightcurves when using the simple flux calculation method, the red dashed line shows the CCF when using the \ion{Mg}{2} lightcurve determined through Fe template fitting.  The low peak value of the CCF indicates no significant correlation or lag.  The dotted line shows the auto-correlation function of the UVW2 lightcurve.}
\label{fig:ccf}
\end{figure}

We use the standard cross correlation analysis techniques to determine if there is a time lag between the UV continuum and the \ion{Mg}{2} line.  Since the UVW2 lightcurve is significantly longer than the 2950--3150\,\AA\ lightcurve, we use the UVW2 lightcurve to search for a lag.  We calculate the cross correlation function (CCF) between the \ion{Mg}{2} and UVW2 lightcurves, using the linear interpolation method as described by \citet{white94}.  The CCFs are shown in Figure~\ref{fig:ccf}  when using the \ion{Mg}{2} lightcurves determined both from the simple and Fe template fitting methods.  Regardless of the \ion{Mg}{2} lightcurve used there are two peaks in the CCF with one at approximately 20 days and the other at approximately 70 days.  However, the CCF does not peak at a high value (the peak is $<0.5$ for both \ion{Mg}{2} lightcurves), indicating that the UVW2 and \ion{Mg}{2} lightcurves are not strongly correlated.  

We use the standard Monte Carlo flux randomization/random subset sampling method as implemented by \citet{petersonetal04} in order to generate 10,000 pairs of lightcurves, and we determine the peak and centroid value of the CCF for each pair.  We use the mean value of the distribution centroid values as our best value for lag centoid, $\tau_{\rm cent}$, though we note that the centroid distribution is also double peaked, like the CCF.  We find  $\tau_{\rm cent} = 34^{+34}_{-24}$ days  when using the \ion{Mg}{2} lightcurve from the simple method and $\tau_{\rm cent} = 13^{+11}_{-12}$ days when using the Fe template method lightcurve, again indicating there is no significant non-zero lag.   Note that the CCFs from both methods peak at around 20 days, and the difference in centroid lags comes from the stronger secondary peak at 70 days in the CCF from the simple method. We also note that the peak in the CCF that can be seen at around 70 days in Figure~\ref{fig:ccf} is narrower than the auto-correlation function (ACF) of the UVW2 lightcurve, and thus likely not real.  If the line lightcurve is formed in response to changes in the continuum lightcurve then the CCF will be the ACF convolved with the transfer function, and so should be broader (not narrower) than the ACF.  Furthermore, the ACF of the \ion{Mg}{2} lightcurve is very narrow, with a secondary peak at 50 days.  As noted when discussing the \ion{Fe}{2} lightcurve, we also tried binning up the spectra in time by up to 5 epochs, however, this also led to no lag detection in \ion{Mg}{2}.

We also determine whether there is any significant lag between the UVW2 and grism continuum lightcurves.  Since the grism lightcurves are close to monotonic, we take out the long term trends in the lightcurves by fitting a low-order polynomial, and subtracting it from the data.  Such detrending is shown to improve lag measurements \citep{welsh99}, and since we expect the interband lags to be short removing the long-term trends reduces aliasing effects.  The lags measured relative to the UVW2 are $\tau_{\rm cent} = -0.9^{+0.8}_{-0.9}$ days, $\tau_{\rm cent} = 1.1^{+0.7}_{-0.5}$ days and $\tau_{\rm cent} = 0.3^{+0.9}_{-1.9}$ days, for the 2015--2215\,\AA, 2950--3150\,\AA, and 4430--4625\,\AA\ lightcurves respectively.
 
Wavelength-dependent continuum lags are expected from disk reverberation (thermal reprocessing), where the inner, hotter accretion disk responds to variations in the irradiating flux before the outer, cooler disk \citep[see][for a detailed description]{cackett07}.  \citet{mchardy14} use the photometric lightcurves during the 2013 campaign to provide a measure of these wavelength-dependent lags, and the even higher cadence \swi\ monitoring in 2014 have provided an extremely good measure of these \citep{edelson15}.  While none of our interband lags are significantly non-zero, they are consistent with those expected based on \citet{edelson15}.

\section{Discussion}\label{sec:disc}

We monitored \ngc\ for approximately 170 days with the U grism onboard \swi\ during 2013, obtaining low-resolution UV spectra approximately every other day during the monitoring campaign.  By combining these spectra with photometric monitoring before and during our grism campaign, we attempted to compare the variability of the continuum and \ion{Mg}{2} emission line.  The UV continuum showed significant variability over the campaign, though the flux changes were almost a monotonic increase over the period of grism monitoring.  The \ion{Mg}{2} emission line, however, was not significantly variable, and hence there is no plausible lag between the continuum and emission line, preventing a reverberation mapping mass estimate from the \ion{Mg}{2} line alone.  

Even though not significant, it is interesting to note that the peak in the CCF between the UV continuum and \ion{Mg}{2} line at approximately 70 days is consistent with the largest peak in the equivalent CCF for \ion{Mg}{2} from the {\it IUE} monitoring campaign of 1989 \citep{clavel91}, though that peak is also of low significance, and quite probably a result of aliasing.  An important consideration is whether the lack of response to continuum variability is due to  an intrinsic property of the line, the location of the \ion{Mg}{2} line emitting region, or whether the near monotonic increase in continuum flux over the grism monitoring period prevented a clear lag measurement.   \ion{Mg}{2} is a low excitation line, thus is emitted from a region further away from the central AGN than other BLR lines such as H$\beta$ or \ion{C}{4}.  This location further from the AGN could be why we do not see a clear response to continuum variability.  If the inner part of the BLR blocks a clean view of the central engine, the \ion{Mg}{2} emitting region may see only scattered continuum emission.  Certainly there is significant absorbing and obscuring material in the inner region of \ngc\ as evidenced by the large number of absorption lines seen during both 2013 and 2014 {\it HST} observations of \ngc\ \citep{kaastra14,derosa15}.  How absorption affects the \ion{Mg}{2} line which is actually a resonant doublet, and how changes in the absorption affect the line variability are not clear since the grism data are not high enough resolution to detect any narrow absorption lines.  Such issues are known to cause problems in accurately determining the line profile \citep{denney13}.  On the other hand, the fact that the continuum lightcurve during the grism observations shows a near-monotonic increase with no strong peaks or troughs, means that it would be hard to see a clear response from the line.  The photometric monitoring that took place immediately before our grism observations began does show several strong variable features, however the lag would have to be almost 100 days to see the two clear peaks in that lightcurve.

It is also important to consider the photoionization properties of \ion{Mg}{2}. The observed response of a line to continuum fluctuations will depend both on its local responsivity, that is the marginal response of the line to continuum variations, and geometric dilution, i.e. the blurring of the response due to the distribution of delays set by the geometry of the BLR.  Photoionization calculations show the responsivity for \ion{Mg}{2} to be low compared to high ionization lines \citep{goad93, obriengg95, koristagoad00}, meaning that the line should not be expected to respond strongly to changes in continuum flux.  For instance, figure 7 of \citet{goad93} which shows how the responsivity for \ion{Mg}{2} compares to other lines.  This can also be seen in figure 2b of \citet{koristagoad00}, which shows that the EW of \ion{Mg}{2} is generally strongly negatively correlated with the incident ionizing photon flux.  This equates to  generally small values in the local gas responsivity for \ion{Mg}{2}, and also results in large centroids in delay in its 1D transfer function. The latter makes this line's response also susceptible to geometric dilution.  The result is a small response amplitude in \ion{Mg}{2}, compared to the other UV emission lines \citep[see also figure 5a of][]{koristagoad00}. Thus, the lack of variability of \ion{Mg}{2} could be due to an intrinsic property of the line, as suggested by \citet{goad99b} when discussing the lack of \ion{Mg}{2} variability in NGC~3516.

It would be valuable to be able to measure \ion{Mg}{2} reverberation mapping masses directly (as we discussed in section 1).  Unfortunately, the lack of significant lag prevents us from doing this.  Although a direct reverberation mapping mass was not possible, we can still use the mean spectrum in order to obtain a `single-epoch' \ion{Mg}{2} mass estimate to compare with the well-constrained H$\beta$ mass for \ngc.   To get a `single-epoch' mass measurement, we combine the line width ($\sigma = 2351$~km~s$^{-1}$) with the mean continuum luminosity at 3000\,\AA, $\lambda L_\lambda$ (3000\,\AA) $= 5.2\times10^{43}$ erg s$^{-1}$, and use the relation of \citet{mcgill08} to estimate $M_{\rm BH} = 7\times10^7~M_\odot$.  This is slightly larger than, though still in reasonable agreement with, the H$\beta$ reverberation mapping mass for \ngc\ of $5.95\times10^7~M_\odot$ \citep[assuming an $f-$factor;][]{grier13}, and $3.2\times10^7~M_\odot$ \citep[from direct modeling;][]{pancoast14}.

In summary, we do not detect a significant lag between the \ion{Mg}{2} and UV continuum flux in \ngc\ from a $\sim6$ month monitoring campaign with \swi.  However, \swi's ability to perform long-term monitoring of nearby Seyfert 1s and obtain crude UV spectra could lead to the measurement of a \ion{Mg}{2} lag in other bright AGN if the \ion{Mg}{2} line is variable enough.

\acknowledgements
We thank the \swi\ team for their hard work and efforts in successfully scheduled this monitoring campaign.  Thanks also to Paul Kuin for use of his software and significant help and discussion about  \swi/UVOT grism analysis.  We thank Mike Goad and Kirk Korista for insightful discussions on the lack of \ion{Mg}{2} variability and also thank the referee for suggesting \ion{Fe}{2} template fitting.  EMC gratefully acknowledges support provided by NASA through grant NNX14AC23G to Wayne State University.  MCB gratefully acknowledges support from the NSF through CAREER grant AST-1253702 to Georgia State University.  BMP gratefully acknowledges support from the NSF through grant AST-1008882 to The Ohio State University. MV gratefully acknowledges support from the Danish Council for Independent Research via grant no. DFF 4002-00275.

\bibliographystyle{apj}
\bibliography{apj-jour,agn}

\end{document}